\documentclass{emulateapj}
\journalinfo{\@Accepted April 30, 2012 for publication at the Astrophysical Journal; Submitted January 19, 2012.}

\shorttitle{Disk assembly - TF Relation to $z$$\sim$1.7}
\shortauthors{Miller et al.}

\begin{document}

\title{The Assembly History of Disk Galaxies: II. \\
Probing the Emerging Tully-Fisher Relation during  1$<$z$<$1.7}

\author{
Sarah H. Miller\altaffilmark{1,2},
Richard S. Ellis\altaffilmark{2},
Mark Sullivan\altaffilmark{1},
Kevin Bundy\altaffilmark{3},
Andrew B. Newman\altaffilmark{2}, \&
Tommaso Treu\altaffilmark{4}
}

\email{s.miller1@physics.ox.ac.uk}

\altaffiltext{1}{Oxford Astrophysics, Oxford, OX1 3RH, UK}
\altaffiltext{2}{California Institute of Technology, Pasadena, CA 91125}
\altaffiltext{3}{Institute for the Physics and Mathematics of the Universe (IPMU), University of Tokyo, Japan}
\altaffiltext{4}{Physics Department, University of California, Santa Barbara, CA, 93106}

\begin{abstract}

Through extended integrations using the recently-installed deep depletion CCD on the red arm of the Keck I Low Resolution Imaging Spectrograph,  we present new measurements of the resolved spectra of 70 morphologically-selected star-forming galaxies with $i_{AB}<$~24.1 in the redshift range $1 \lesssim z  < 1.7$. Using the formalism introduced in Paper I of this series and available HST ACS images, we successfully recover rotation curves using the extended emission line distribution of [O II] 3727 \AA\  to 2.2 times the disk scale radius for a sample of 42 galaxies. Combining these measures with stellar masses derived from HST and ground-based near-infrared photometry enables us to construct the stellar mass Tully-Fisher relation in the time interval between the well-constructed relation defined at $z$$\simeq$1 in Paper I and the growing body of resolved dynamics probed with integral field unit spectrographs at $z$$>$2. Remarkably, we find a well-defined Tully-Fisher relation with up to 60\% increase in scatter and zero-point shift constraint of $\Delta M_{\ast}$ = 0.02$\pm$0.02 dex since $z\sim1.7$, compared to the local relation. Although our sample is incomplete in terms of either a fixed stellar mass or star formation rate limit, we discuss the implications that typical star-forming disk galaxies evolve to arrive on a well-defined Tully-Fisher relation within a surprisingly short period of cosmic history.

\end{abstract}
\keywords{galaxies: evolution --- galaxies: fundamental parameters --- galaxies: kinematics and dynamics --- galaxies: spiral}

\section{Introduction}

Excellent progress has been made over the past 15 years in observationally charting the history of cosmic star formation \citep{hopkin2006,ellis2008}. Attention has now moved to interpreting the stellar mass assembly history of galaxies \citep[e.g.,][]{bundy2006} in a framework whereby cold dark matter halos merge under gravitational instability \citep{white1978}. Combining dynamical and stellar masses for selected galaxies at various look-back times provides insight to the relative contributions of dark and baryonic matter through time. For the most abundant population of star forming galaxies, the Tully-Fisher relation, which probes the relationship between luminosity and rotational velocity, is the most effective tool in assessing the angular momentum development of disks as they assemble in dark matter halos \citep{fallef1980,blumen1984,momaow1998,navarr2000}. Of particular value is the stellar mass Tully-Fisher relation ($M_{\ast}$-TFR), which is tighter than any scaling relation confined to a given luminosity band \citep{bellde2001} making it highly sensitive to self-similar disk growth through time.

In Paper I of this series \citep{miller2011} we demonstrated, using extended integrating times with DEIMOS on the Keck 2 telescope together with an optimal technique for extracting kinematic rotation curves to a location corresponding to 2.2 disk scale radii (2.2$r_s$ or $r_{2.2}$), a stellar mass TFR at $z\simeq1$ that has virtually the same intrinsic scatter as that observed locally. This indicates that most disk galaxies at LMC-mass or higher are dynamically mature by $z\simeq1$, in contrast to the conclusions of earlier studies based on shorter exposure times or other velocity extraction methods \citep{consel2005, flores2006, kassin2007}. Moreover, the zero point shift in the relationship from $z\simeq1$ to $z\simeq0.2$ is marginal ($\Delta M_{\ast} \sim 0.04 \pm 0.07$ dex), which we found was due to an evolving mix of the baryonic components that likely dominate within $r_{2.2}$. While the exact fraction of dynamically mature disk galaxies at a given redshift is debated, the results of Paper I for disks at $z\simeq1$ contrasts markedly with the dispersion-dominated dynamics of massive, star-forming galaxies observed 2-3 Gyr earlier at $z>2$ \citep{genzel2006,forste2009,jones2010}. In addition to a large increase in scatter,  \citet{cresci2009} show the $z\sim2.2$ $M_{\ast}$-TFR has an offset of $\Delta M_{\ast} \sim 0.41 \pm 0.11$ dex from that found in Paper I. The implication is that star-forming galaxies very rapidly establish their present dynamical state during the intervening period.  Thus, resolved dynamical data in the redshift interval $1<z<2$ is needed to make progress, and large samples will be valuable to understand the demographic trends. 

\begin{figure}
\includegraphics[width=3.4in]{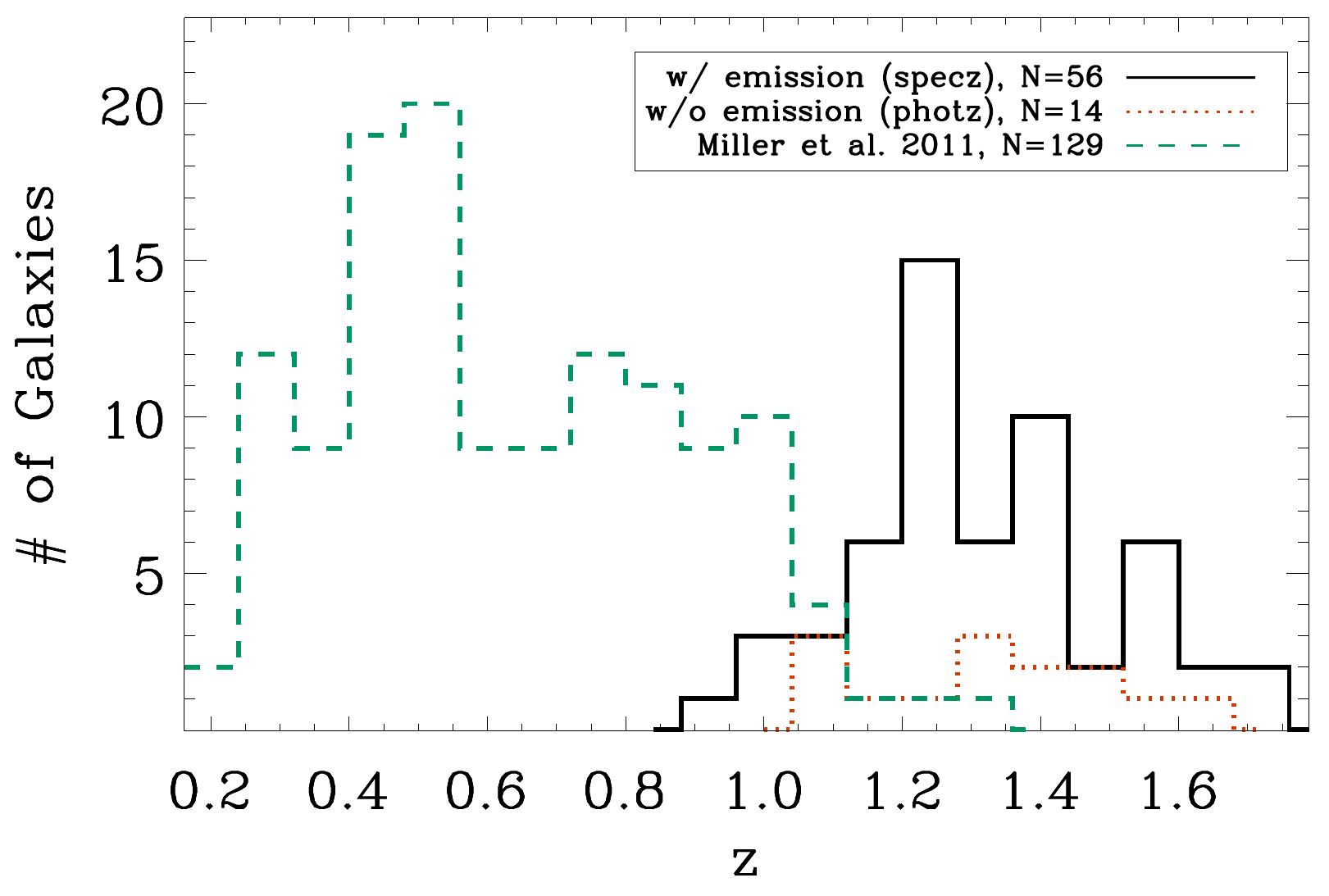}
\caption{The redshift distribution of 70 galaxies in our new sample compared to that in Paper I. Line emission was detected in most of the $z>1$ galaxies targeted as shown in solid black. The dotted orange line shows the distribution for those galaxies with no detected emission.}
\label{fig:zhist}
\end{figure}

There are several observational challenges in pursuing the $M_{\ast}$-TFR beyond $z\simeq1$. Multi-object capabilities are essential in creating a large sample to address questions such as those posed above, but these capabilities are only currently available at optical wavelengths. The emission line of choice, [O II] 3727 \AA~moves to a wavelength region of low detector sensitivity, and the physical scale being probed is comparable to just a few times the average seeing (a diameter of 5$r_s$ at $z\simeq1.6$ corresponds to $\simeq$2.5 arcsec). This second paper in our series was motivated by the installation of a deep depletion, red-sensitive CCD on the red arm of the Low Resolution Imaging Spectrograph \citep[LRIS, ][]{oke1995} which has significantly improved the efficiency in the wavelength interval 8,000~\AA~to~10,000~\AA~corresponding to [O II] in the redshift interval $1.0\lesssim z<1.7$ (Fig.~\ref{fig:zhist}). This allows us to extend our earlier work,  to try to understand the transition from the dispersion-dominated systems observed above $z$$\simeq$2, and to evaluate the nature of the $M_{\ast}$-TFR 2 Gyr earlier than that probed in Paper I. We present the results based on a sample of 70 morphologically-selected star-forming galaxies for which we have obtained the necessary extended integration times, following the techniques developed in Paper I.


The plan of the paper is as follows: in \S 2, we describe the sample selection criteria, the LRIS data and the HST Advanced Camera for Surveys (ACS) resolved photometry and stellar mass estimates; \S 3 explains our technique in determining the fiducial rotation velocities, as well as various tests of the chosen method; in \S 4, we present our results of the TF relation from  $1.0 \lesssim z < 1.7$, and in \S 5 we discuss our results in light of current disk assembly theory and pioneering observations to even higher redshift. Throughout the paper we adopt a $\Omega_{\Lambda}$ = 0.7, $\Omega_{m}$ = 0.3, $H_0$ = 70 km sec$^{-1}$ Mpc$^{-1}$
cosmology. All magnitudes refer to the AB system.

\section{Data}

We selected our sample of spectroscopic targets in a similar manner to that in Paper I so as to span a broad mass and luminosity range. A more detailed look at sample representativeness can be found in the Appendix, but the salient points are summarized in the following section. The necessity of HST ACS imaging for accurate disk size measurements led us to focus on five fields with multi-wavelength coverage for selecting targets: (1) the Great Observatories Origins Deep Survey (GOODS) North and (2) GOODS South fields \citep{giaval2004}, (3) the Small Selected Area 22 (SSA22) field \citep{lilly1991,chapma2004, abraha2007}, (4) the Extended Groth Strip (EGS) field \citep{davis2005}, and (5) the Cosmic Evolution Survey (COSMOS) field \citep{scovil2007}. A summary of masks, targets, spectroscopic observations and available space and ground-based imaging in each field can be found in Table \ref{tab:samp}. 

\begin{deluxetable*}{lllllllll}
\tablewidth{0pc} 
\tablecaption{Summary of LRIS observations and data} 
\startdata 
\tablehead{ \colhead{{\tiny FIELD}} & \colhead{$N_{msk}$\tablenotemark{a}}  & \colhead{$N_{obj}$\tablenotemark{b}}   & \colhead{$t_{exp}$\tablenotemark{c}} & \colhead{{\tiny DATES OF OBS} \tablenotemark{d}} & \colhead{{\tiny SEEING}}  & \colhead{{\tiny HST Filters}\tablenotemark{e}}  & \colhead{{\tiny Ground IR Phot}\tablenotemark{f}} & \colhead{{\tiny IR Catalog}\tablenotemark{g}}}

{\tiny EGS} & 2 & 11(3,3) & 30.0~ks & {\tiny 26--28 Jun 2009} & 1\farcs05 & \emph{F606W, F814W} & {\tiny Palomar} ($JK_s$) & \citet{bundy2006} \\
& &  12(2,2) & 14.4~ks & ''  & 0\farcs93 &    &   & \\ 
{\tiny SSA22} & 2 & 13(3,3) & 22.5~ks & '' & 0\farcs78 & \emph{F814W} & {\tiny UH~2.2m} ($JHK_s$) & \citet{capak2004} \\
& &  12(2,1) & 12.0~ks & ''  & 0\farcs89 &    &   & \\ 
{\tiny GOODS N} & 1 & 8(1,3) & 34.8~ks & {\tiny 5--6 Apr 2010} & 0\farcs86 & \emph{F435W, F606W}  & {\tiny MOIRCS} ($K_s$) & \citet{bundy2009} \\ & & & & & &  \emph{F775W, F850LP} &  on {\tiny Subaru} & \\ 
{\tiny COSMOS} & 1 & 9(2,1) & 28.8~ks & {\tiny 7 \& 10 Jan 2011}& 0\farcs82 &  \emph{F814W} & {\tiny WIRCam} ($K_s$)  & \citet{mccrac2010} \\
& & & & & &  & on {\tiny CFHT} & \\ 
{\tiny GOODS S} & 1 & 5(1,1) & 25.2~ks & {\tiny 28 Feb 2011} &  0\farcs98 & \emph{F435W, F606W} & {\tiny ISAAC} ($K_s$) & \citet{retzla2010} \\
& & & & & &  \emph{F775W, F850LP} &  on {\tiny ESO VLT}  & \\ 
\enddata
\tablenotetext{a}{number of masks per field}
\tablenotetext{b}{number per mask and those of which are (compact, passive), see \S\ref{sec:mod}}
\tablenotetext{c}{total integrated exposure time per mask}
\tablenotetext{d}{dates observations were taken at Keck I}
\tablenotetext{e}{filters of HST ACS imaging available for size measures}
\tablenotetext{f}{Ground-based infra-red photometry available for SED fitting, in addition to further optical filters (not shown)}
\tablenotetext{g}{Reference to photometry matching catalog}
\label{tab:samp} 
\end{deluxetable*}

In selecting our targets we applied an apparent magnitude limit of $i_{AB}<25$ and selected sources in the redshift range $1.0<z<1.7$ (using photometric redshifts when a spectroscopic redshift was unavailable).  This led to a sample of 50 targets per LRIS pointing. Potential targets were then visually inspected by at least two of us (RSE, KB, SHM) on the ACS images and potential targets with compact morphology, without diffuse extension, were given low priority but not excluded.\footnote{In considering the effect of excluding
	unresolved or compact sources, it is important
	to recognize that the dynamical measurements
	for spatially unresolved galaxies will likely add
	to the observed scatter through their measurement 
	uncertainties rather than because those sources have
	an intrinsically larger scatter around the
	relationship.} From a prioritized list of typically 25 suitable targets per LRIS pointing, on average $\sim$10 of these were included per mask due to the requirements  of both suitably-oriented slits and simultaneous accommodation of unrelated targets for other science campaigns.  In designing multi-slit masks for LRIS, we aligned 1 \arcsec~slitlets to the major axes of our targets as fit by \textsc{SExtractor} \citep{bertin1996}.  The position angle (PA) of the mask was then selected to minimize slitlet tilts while maximizing target count, and we always ensure that target slitlets are within $\pm$30$^\circ$ of the mask PA.  We used the 600~l mm$^{-1}$ grating blazed at 10000 \AA, providing a instrumental velocity resolution of  58~km~s${}^{-1}$ at 9000~\AA\ thereby exploiting the sensitive region of the newly-installed CCD. The spectroscopic LRIS data were reduced by one of us (AN), using techniques described in detail by \citet{newman2010} with the code developed by \citet{kelson2003}, in which sky modeling and subtraction is carried out with sub-pixel sampling of the background. The seeing, which varied from 0\farcs78 to 1\farcs05, was measured by taking the weighted average of the FWHM of the alignment stars on each mask.

\begin{figure*}
\begin{center}
\includegraphics[height=3in]{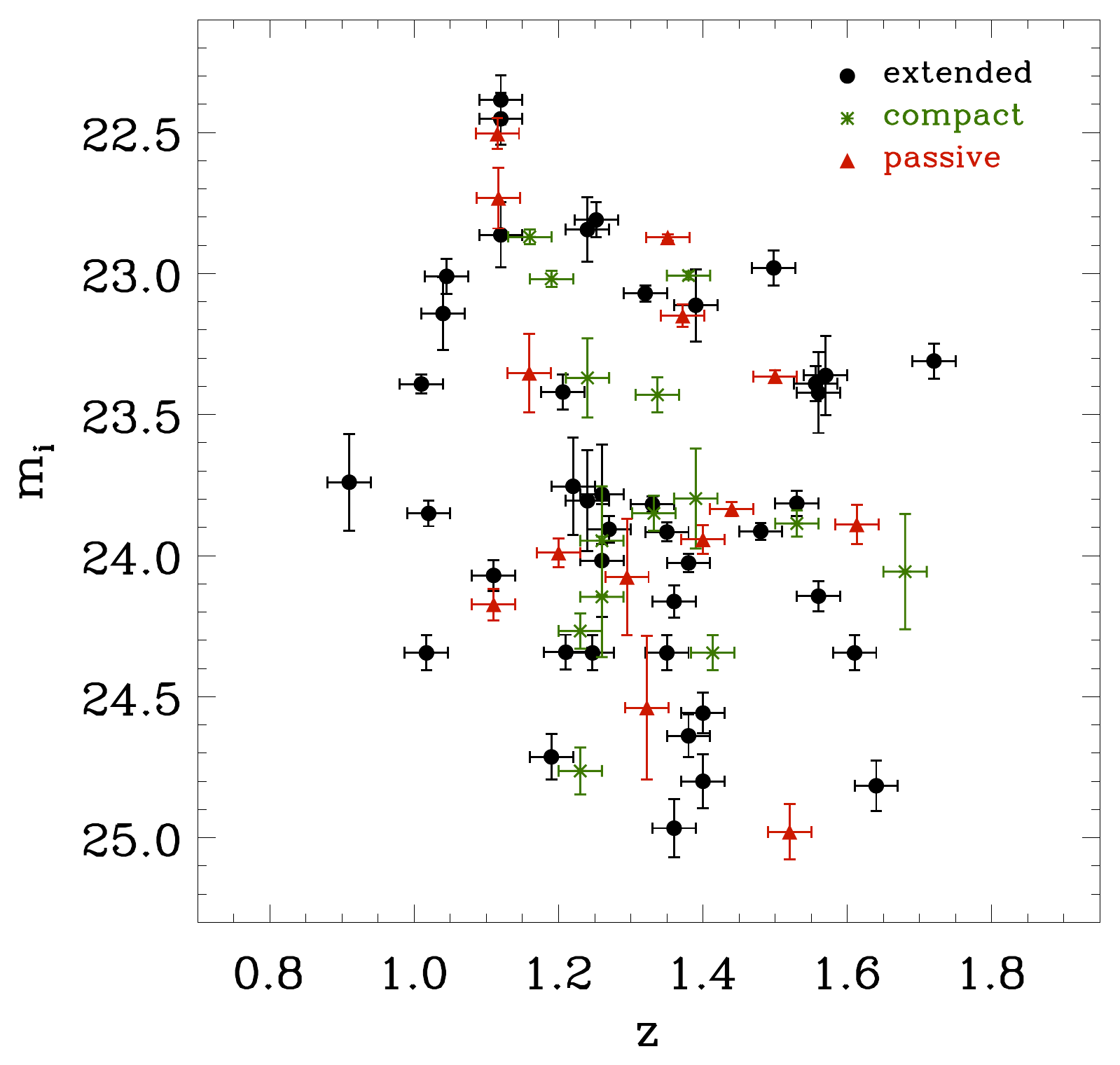}
\includegraphics[height=3in]{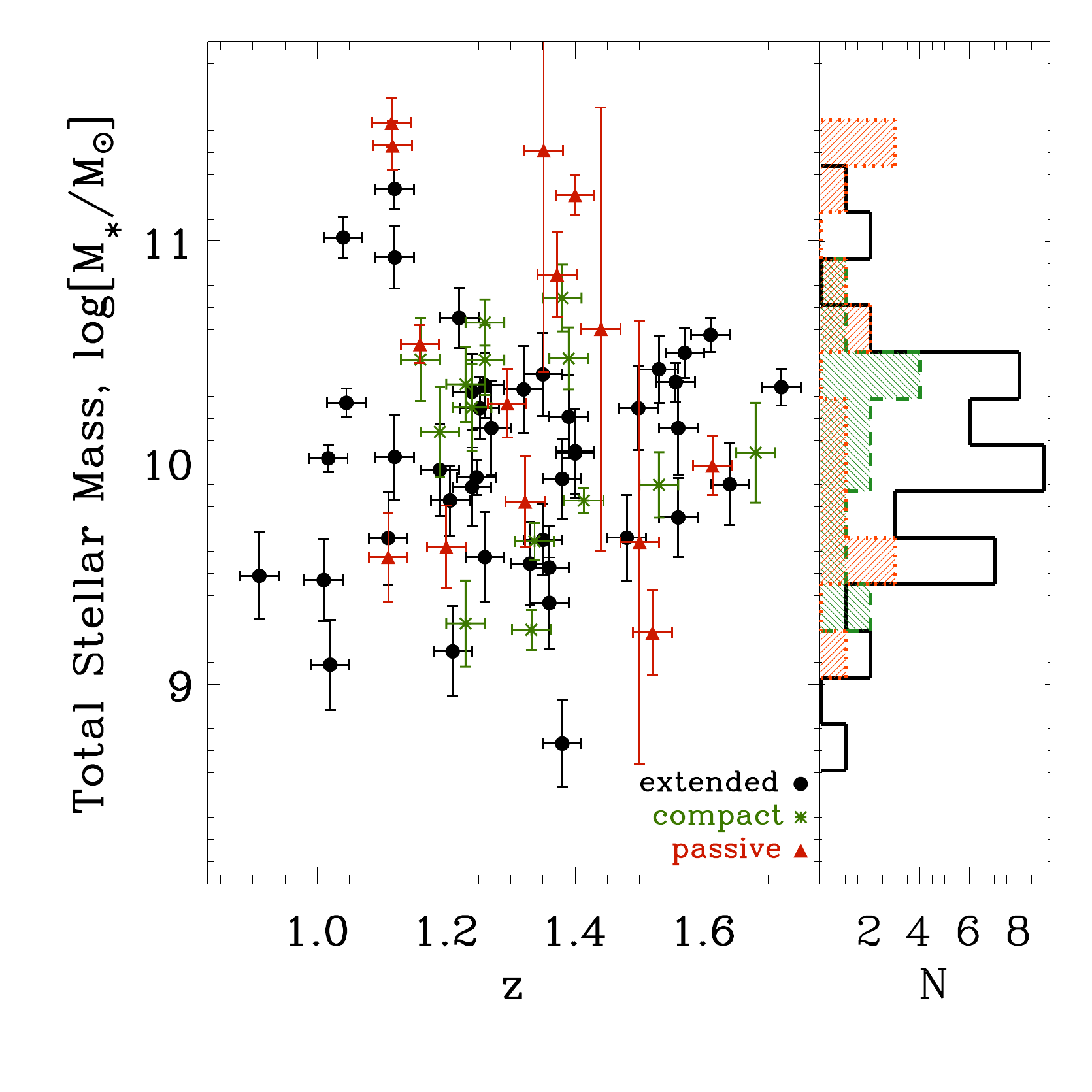}
\includegraphics[height=2.3in]{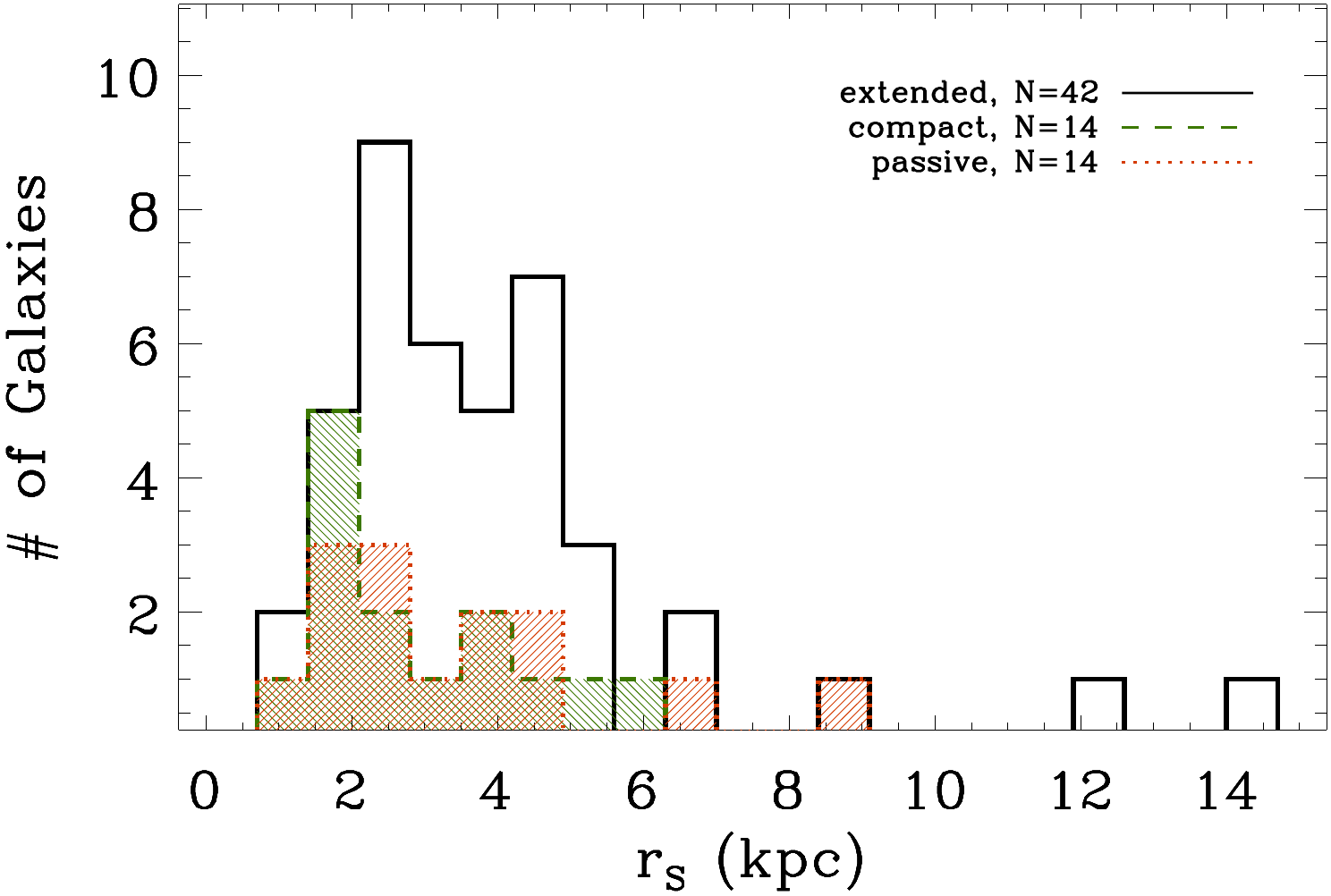}
\includegraphics[height=2.3in]{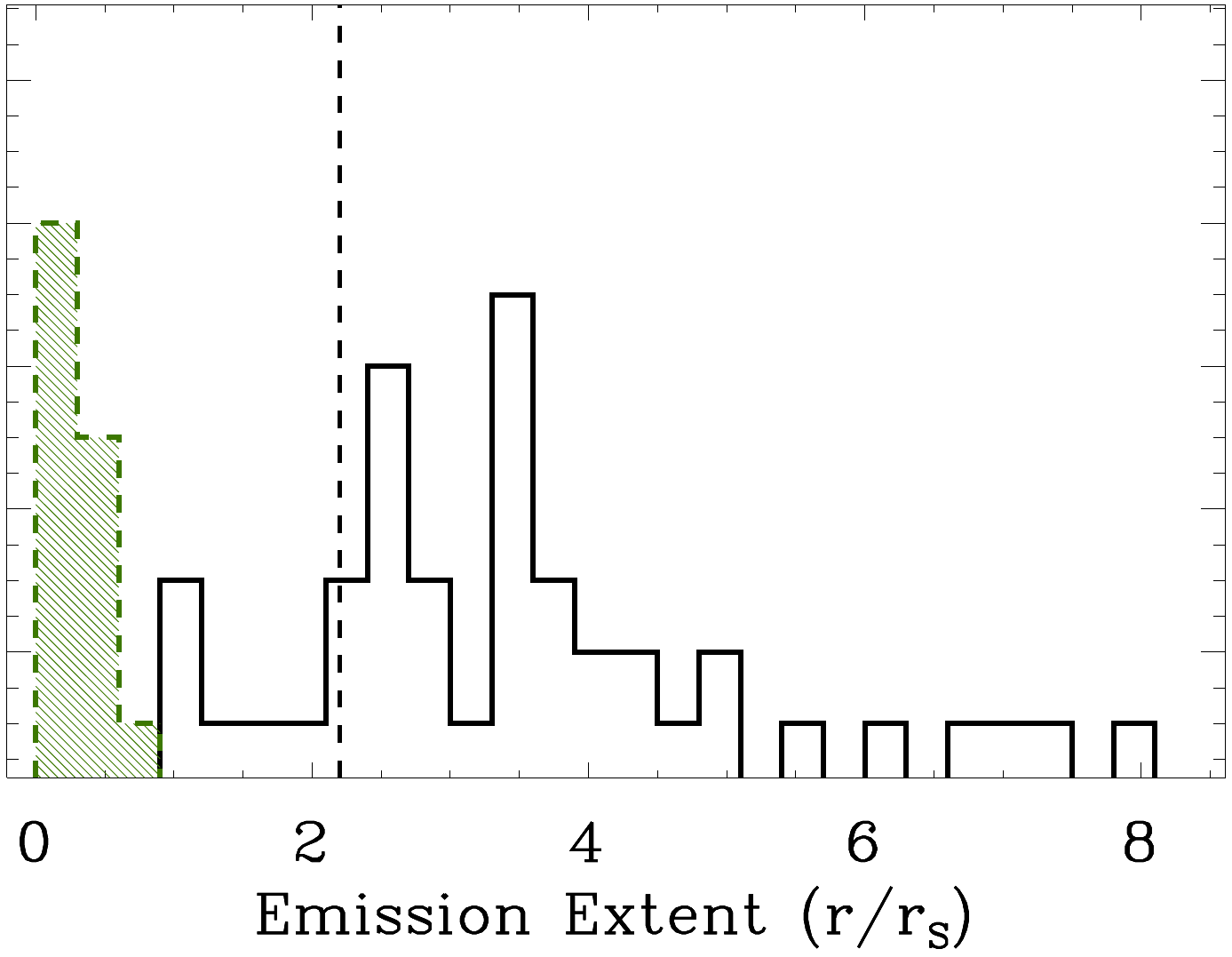}

   \caption{Distributions of redshift, apparent $i$ magnitude, stellar mass, exponential disk scale radius ($r_s$), and the extent of [O II] line emission for our sample (measured in terms of the disk scale radius $r_s$). The sample is partitioned into the three spectral categories discussed in the text: extended line emission (black line), compact line emission (green dashed line), which does not extend past the seeing or 1 scale radius, and passive sources (orange dotted line). The fiducial radius of 2.2 times the disk scale radius $r_s$ is plotted with the vertical black dashed line. See \S\ref{sec:dyn} for further details.}
   \label{fig:hists}
   \end{center}
\end{figure*}

\subsection{Photometric Components} \label{sec:phot}

With the existing HST ACS data (see Table \ref{tab:samp} for availability per field), 
we conduct our measurements of disk radii and  inclination of each galaxy 
using the reddest available filter (F814W or F850LP). We ensure that light from the bulge 
component does not confuse the fitted disk component by conducting 
a bulge-disk decomposition, where appropriate, similar to our procedure in Paper I.

The HST images were analyzed for photometric components using the \textsc{galfit3}
\citep{peng2010} least squares elliptical-fitting method described in detail in Paper I
to which the reader is referred. Briefly, we fit an exponential disk component plus a de Vaucouleurs'
bulge profile to every galaxy. Those galaxies which yielded unphysical
solutions were re-fit with a single Sersic profile component, where
the Sersic index ($n$) was allowed to vary. Such cases generally
represent disk galaxies which are bulgeless and/or more clumpy and
irregular than regular well-formed spirals. For the present sample, $\sim$63\% of our galaxies
were fit using a 1-component $n$-varying Sersic profile fit, and
$\sim$37\% were adequately fit with a two-component bulge and disk
solution. This mix provides a valuable indication of the morphological
distribution of our sample, suggesting less than half are
well-formed disks. For comparison, this mix was 60:40 in Paper I.

We ran \textsc{galfit} using HST data of the reddest available filters (noted in Table \ref{tab:samp}). 
Where imaging from more than one filter is available, we run
 \textsc{galfit} on each and see $<$10\% difference between bands. 
 In our comparisons between the F435W filter in Paper I and the redder filters, we
see no systematic offset, but typically a $<$5\% increase in the scatter of the radius fit. We assume
an additional uncertainty of this nature exists in the measurements of this dataset since we are sampling rest-frame blue-UV light in the majority of the sample.
 To achieve convergence on the \textsc{galfit} parameters and to assess 
their errors, we run a similar Monte Carlo analysis to that in Paper I. We found 
the parameter output distributions 
were much narrower than the input distributions, suggesting
convergence.  Final parameter uncertainties from 
the Monte Carlo distributions are better than 10\% on average, and the
additional uncertainties are added in quadrature to the observational error
and formal fitting errors.

\subsection{Stellar masses} \label{sec:masses}

Our method of estimating stellar mass in our galaxies follows the work initially presented 
in \citet{bundy2005}, followed by the analysis presented in \citet{bundy2009}. Further details 
can be found in those papers.

Briefly, stellar masses are derived using a matched catalog of multi-band ACS, available optical 
and ground-based 
infra-red photometry (see Table \ref{tab:samp} for ACS and infra-red availability per field). A Bayesian code fits the 
spectral energy distribution (SED) derived from 2\arcsec~ACS, and ground based optical and infra-red photometry, adopting 
the best spectroscopic redshift and this SED is compared to a grid of \citet{bruzua2003} models 
that span a range of metallicities, star formation histories, ages and dust content. The stellar mass 
is calculated from the derived $K$-band mass/light ratio assuming a \citet{chabri2003} initial mass 
function (IMF). The probability for each fit is marginalized over the grid of models giving a stellar 
mass posterior distribution function, the median of which is the catalogued value. At the magnitudes 
probed in this survey, the uncertainties inferred from the median 68\% equivalent width of these posterior functions is 0.174 dex. 

In order to determine the systematic uncertainties arising from combining different sets 
of photometry into a single sample, we compare mass estimates for our sample
in SSA22 using different combinations of the available photometry, e.g. the filters
available in EGS. The resulting dispersion in the distribution of stellar mass estimates
ranges per galaxy from 0.005 to 0.1 in dex, with a median standard deviation of
0.052, which we add in quadrature to each galaxy's SED median posterior width as an additional systematic uncertainty.
The combined stellar mass uncertainty is better than 0.2 dex in over 80\% of our sample.\footnote{This does not include systematic uncertainties regarding our incomplete knowledge of stellar populations in the various types of galaxies present in this sample at these redshifts--- uncertainty which affects all present studies of this nature.}
As in Paper I we use the stellar mass enclosed within 2.2$r_s$, as scaled
from the reddest available HST broad-band flux, in order to compare like velocity with like mass. The average reduction in 
stellar mass at $r_{2.2}$ is -0.187 dex compared to the total stellar mass estimates to the Kron 
radius (Kron 1980). For the present sample, we successfully probe down to a 
 minimum total stellar mass of log $M_{\ast}/M_{\odot}$ of 8.73 dex, with a mean of 10.04 dex.

\section{Dynamical analysis} \label{sec:dyn}

\begin{figure*}
   \centering
     \includegraphics[width=6in]{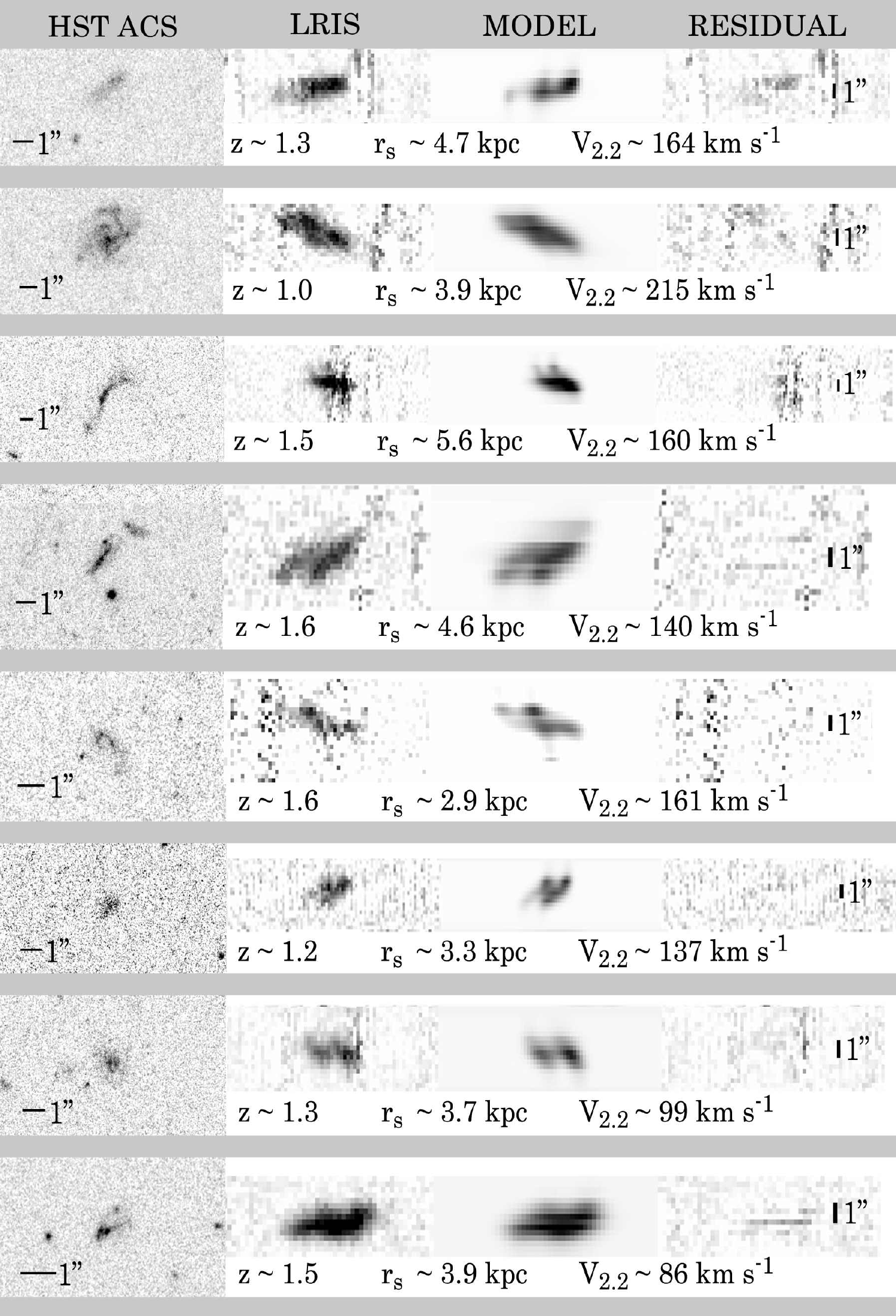} 
   \caption{Examples of our kinematic analysis: (from left to right) the HST image of the galaxy in ACS \emph{F850LP} or \emph{F814W} filter; the 2-D [OII] data in the reduced LRIS spectra; the modeled emission trace using an arctan function blurred by both the measured seeing and dispersion (\S\ref{sec:mod}); the residual obtained by comparing the latter two signals. }
   \label{fig:ex}
\end{figure*}

Modeling rotation curves beyond $z\sim1$ is a challenging endeavor in many respects. In addition to the shift of the diagnostic [O II] 3727 \AA\ emission line into the sky-dominated red spectral region, the more distant galaxies are smaller in angular scale making resolved measurements challenging from the ground without the use of adaptive optics (AO). Prior to the availability of near infra-red (NIR) multi-IFU instruments (e.g., KMOS on the VLT), multi-slit spectrographs retain several advantages for delivering kinematics for large samples of galaxies over a wide redshift range. Only a fraction of the presently-available IFU datasets have been observed with AO, and the typical wider wavelength ranges and multi-plexing capabilities of spectrographs such as DEIMOS, LRIS, and the soon-to-be-commissioned MOSFIRE on Keck enable extended exposures on survey sources without a penalty in spectral or spatial resolution. Improved multiplexing of IFUs will mitigate this current deficiency as 2-D spectroscopy offers clear advantages in optimal fitting of complex velocity fields \citep{puech2008,jones2010,genzel2011,forste2011,contin2011}.  For now, multi-slit spectrographs retain a key advantage in survey efficiency. Here we exploit the improved red-sensitivity of
a newly-installed LRIS deep depletion CCD from Lawrence Berkeley National Laboratories \citep{rockos2010} to secure resolved dynamics for 70 galaxies in the redshift interval $1<z<1.7$ with a lower average stellar 
mass range than the many extant IFU samples. 

We discuss our motivation in Paper I  to determine the rotational velocity at a fixed fiducial radius, and why we choose this to be $r_{2.2}$, or 2.2 times the scale radius ($2.2\times r_s$) of the galaxy, as measured from the broad band HST optical filter. A maximum measured velocity, $V_{max}$, is highly dependent on the inconsistent distribution of emitting gas in a given galaxy for studies based on nebular emission lines such as ours. So it is important instead to base a measurement for the TF relation at a consistent location in the disk galaxy such as $V_{2.2}$, the velocity found at $r_{2.2}$ (further justified in Paper I, and references therein). In the following we will briefly review our methodology for extracting rotation curves and determining the velocity, $V_{2.2}$ at this fiducial radius, as well as testing the uncertainties given the increased distance of our sample c.f. the Paper I sample.

\subsection{Rotation Curve Modeling} \label{sec:mod}

As in Paper I we adopt the empirically-motivated arctan function as the starting point of our modeling procedure, viz:
\begin{equation}
V = V_{0} + \frac{2}{\pi} V_{a} \arctan(\frac{r - r_0}{r_t}),
\label{eq:arctan}
\end{equation}
where $V_0$ is the central or systematic velocity, $r_0$ is the dynamic centre, $V_{a}$ is the asymptotic velocity, and $r_t$ is the turnover radius, which is the transitional point between the rising and flattening part of the rotation curve \citep{courte1997, willick1999}.  

Paramount to modeling emission lines for rotation curves is that the emission being modeled is adequately resolved. There are two situations in which this may not occur: (1) the angular extent of the galaxy is insufficient compared to the seeing and spatial sampling of the instrument, and (2) where the galaxy may
be sufficiently extended in broad band imaging but the spectroscopic emission remains unresolved. Our morphological selection has effectively de-selected
the first category but we must now deal with the second.

Ideally, the line emission must extend beyond the seeing disk. To robustly model the dynamics with the arctan function we have generally found that this corresponds to a simple criterion that the line emission must extend to at least one disk scale radius ($r_s$), as measured by the broad-band photometry. By testing our modeling code with simulated data we found in Paper I that too many possible models are consistent with the data if this simple criterion regarding the emission line extent is not met. By progressively removing flux from the outer disk emission in the current LRIS data, we have confirmed this result from Paper I. We thus applied this criterion to define that subset of our sample for which emission line data can be appropriately modeled. Studies which do attempt to dynamically model unresolved kinematics will very likely introduce scatter into their scaling relations. Unresolved rotational velocity gradients, in both IFU and multi-slit datasets, will propagate through the instrument as increased dispersion, degenerate with the intrinsic velocity dispersion in the optical and NIR -emitting gas of the galaxy.  Especially if the unresolved velocity gradient of the inner solid body rotation is the only part of the galaxy with sufficient signal-to-noise, fitting a model such as the arctan function in Eq.\ref{eq:arctan} will result in poorly constrained extrapolated velocities toward the outer disk, e.g. any velocity wishing to sample the flat part of the rotation curve for a Tully-Fisher study \citep{courte1997,miller2011}.

Of the 70 spectra discussed earlier, 60\% (42) show emission which is sufficiently extended according to the above criterion from which a rotation curve can be modeled. Of the remainder,  20\% (14) have detectable line emission but it is too compact, i.e. does not extend above $r_s$, and 20\% (14) appear to be passive without emission. This 60/20/20 composition of extended, compact, and passive spectral emission respectively, is similar to the equivalent mix (55/25/20) seen in Paper I, and as before, these three sub-samples are not statistically different in redshift (Fig.~\ref{fig:hists}). For the spectrally compact sample, it appears that 6 out of the 14 galaxies are simply too small to resolve, and the other 8 are well extended in their broad-band optical disk, but are only undergoing significant star formation in their central regions, within 1 disk scale radius. These spectrally compact objects as well as the spectrally passive objects have larger stellar masses by 0.2 dex than disks with extended emission. Spectrally passive disks have only slightly redder $I-K$ colors (by $>1\sigma$) than the distribution width of extended emission galaxies, but show no significant difference in size compared to those with extended emission.

As in Paper I (see for full detailed modeling procedure), we blur the arctan function in both the spatial and spectral dimensions to account for seeing and instrumental dispersion, respectively; these are highly degenerate in many cases. In order to attempt to break this degeneracy we measure the position-dependent dispersion from the LRIS spectra in each 
galaxy, and separately measure the seeing from the alignment  stars in each slit mask. In the model, we implement the dispersion measured from the 
spectrum, reduced by a factor fit to remove additional dispersion of the line from blurring by seeing. We then blur the model by the seeing, and re-normalize the flux across each spatial bin (see Fig.\ref{fig:ex} for examples). The factor to account for the effect of the seeing in the dispersion is fit by the Levenberg-Marquardt least-squares solver, along with the parameters of the underlying arctan function shown in Eq. \ref{eq:arctan}. We then correct the extracted values of $V_{2.2}$ at $r_{2.2}$ for the effects of disk inclination with the minor-to-major axis ratio $b/a$ derived from our \textsc{GALFIT} results. 

For 35 galaxies (half our original sample) our spectra are sufficiently deep that we can robustly follow the line emission to $r_{2.2}$ (after accounting for seeing). For the remaining 7 galaxies with extended line emission, we are forced to infer $V_{2.2}$ by extrapolating beyond the point of maximum emission extent in 
the best fit rotation curve. The error budget for this modeling method is discussed in Paper I, and we discuss further uncertainties regarding this specific application of our spectral models below.

\begin{figure}
   \includegraphics[width=3.45in]{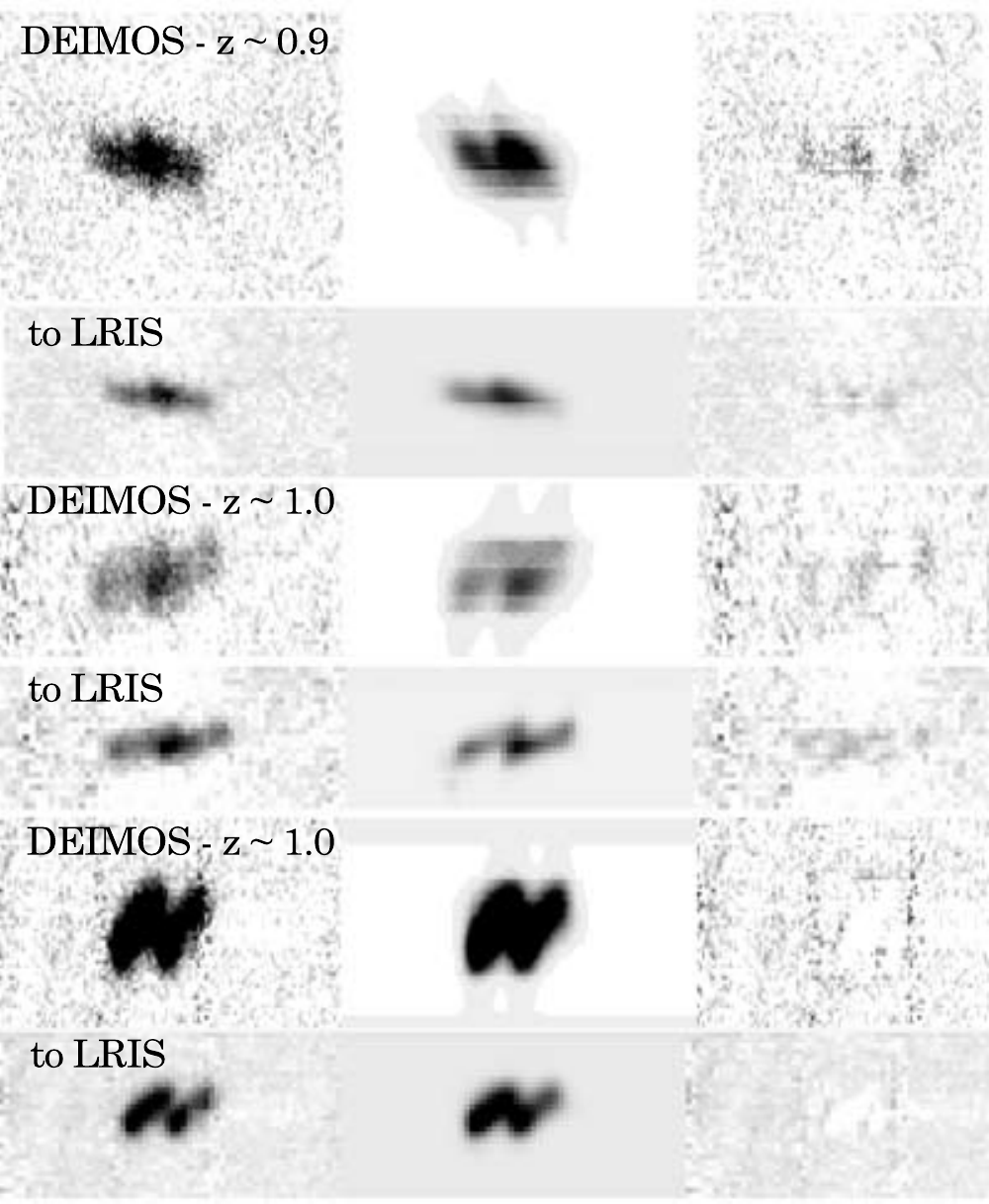}
   \caption{Three galaxies at $z\sim1$ observed with DEIMOS, compared to a simulated observation at the different spatial and spectral resolution of the LRIS instrument directly below. From left to right, the columns show (i) the spectral data, (ii) the best-fit model, and (iii) the residual of the data from the model. }
   \label{fig:relris}
\end{figure}

\begin{figure}
   \includegraphics[width=3.45in]{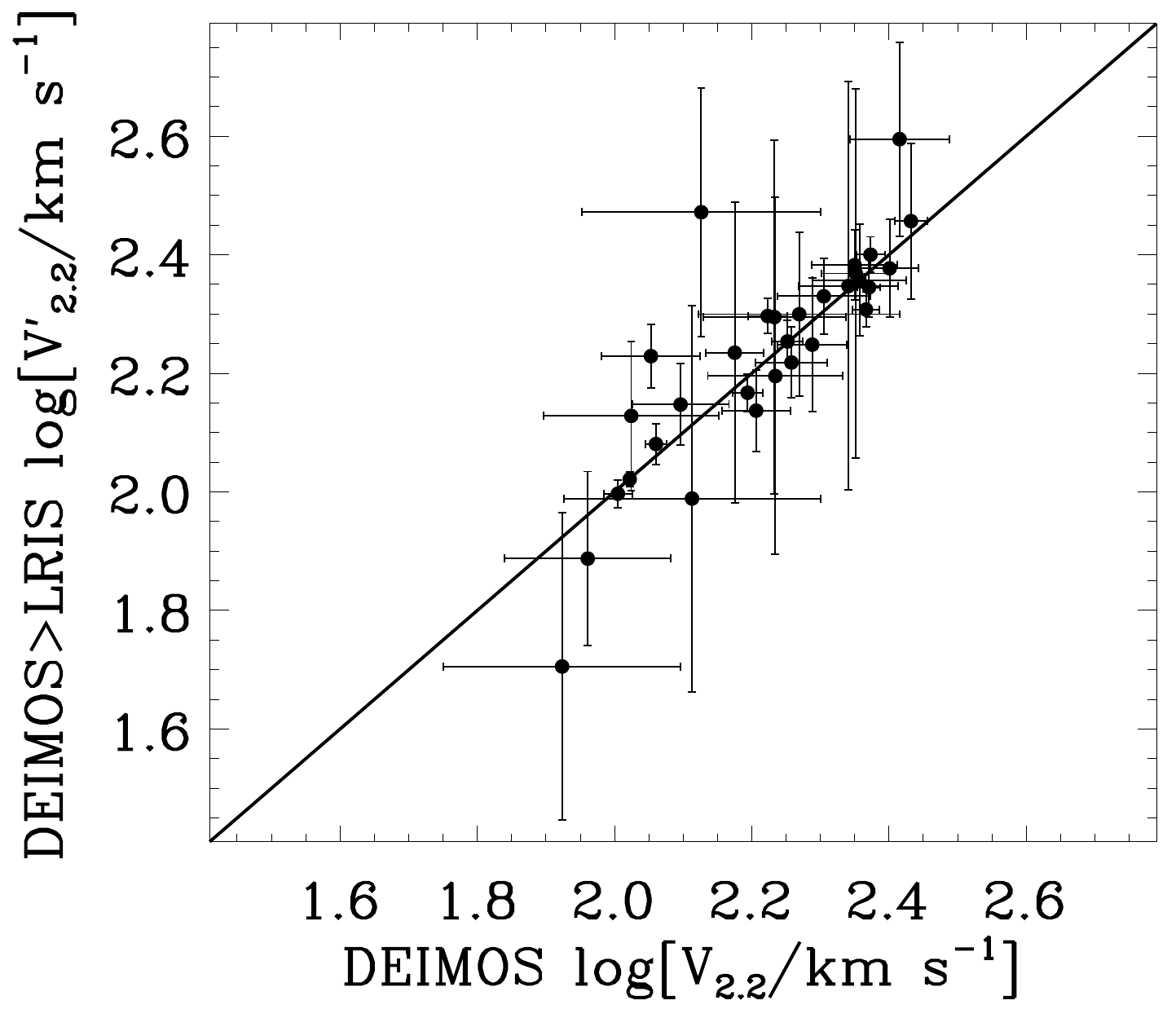}
   \caption{Testing the method: For a subset of galaxies from Paper I in the redshift interval $0.85<z<1.3$ we re-sample their original DEIMOS spectra to the spatial and instrument resolution of LRIS and compare the derived $V_{2.2}$ with the original measures (see text for details).}
   \label{fig:d2l}
\end{figure}

\vspace{0.1in}

\subsection{Testing the Model}

At a redshift $z\simeq1.5$, a typical scale radius of 4 kpc is only $\sim$0.5 arcsec and so clearly seeing and instrumental dispersion blur much of the detail in our LRIS spectra. As it is not always clear visually that there is an intrinsic rotation curve with a characteristic turnover point, we need to demonstrate the reliability and precision of our modeling technique.

To facilitate this, we selected galaxies in the redshift range $0.85<z<1.3$ studied with DEIMOS in Paper I and re-sampled these data to the spatial and spectral resolution of the LRIS data. With the cosmology assumed, the average angular scale of the LRIS galaxies (at $\langle$z$\rangle$=1.31) is 8\% smaller than that for our chosen DEIMOS sample (at $\langle$z$\rangle$=0.98). To account for differences between Paper I and the present survey, we therefore resample for
the different spatial pixel scales (0\farcs1185 with DEIMOS to 0\farcs27 with LRIS) and match the LRIS spectral dispersions (58~km~s${}^{-1}$ at 9000~\AA~with 27.9 km s${}^{-1}$ for each pixel).  We then analyzed the resampled DEIMOS data using the same rotation curve fitting procedure as we use for the LRIS analysis, and the results can be seen in Figs.~\ref{fig:relris}~\&~\ref{fig:d2l}. Encouragingly, when we subtract the recovered $V_{2.2}$ of the resampled DEIMOS data from that of the original data, we find a weighted mean of less than 0.002$\pm$0.009 dex, and the scatter in the relation between the two is 0.098 dex, which is similar to the rms scatter found in the Tully-Fisher relation established locally and in Paper I.  Re-sampling objects at $z\sim1.0$ to the angular diameter distance of that found at $z\sim1.3$ did not make significant changes in the recovered $V_{2.2}$, given this corresponds to only a 8\% reduction in the angular scale.

We also tested the data with models simpler than the arctan-based model. This addresses a possible concern that we might be over-fitting the data given the apparent lack of resolved detail at this somewhat higher redshift.  By assuming a linear fit to the emission at LRIS resolution, the scatter in the resulting estimated $V_{2.2}$ was as large as 1 dex in log[V/km s${}^{-1}$] compared to that of the arctan-based fits at DEIMOS resolution. We also tested whether we could recover $V_{2.2}$ by measuring $dV/dr$ at intervals of $0.1$ scale radii along the disks, but the scatter in the estimated $V_{2.2}$ was on average 0.5 dex, depending on the radial extent of the emission.

\section{Results}\label{sec:results}

\begin{figure}
   \includegraphics[width=3.45in]{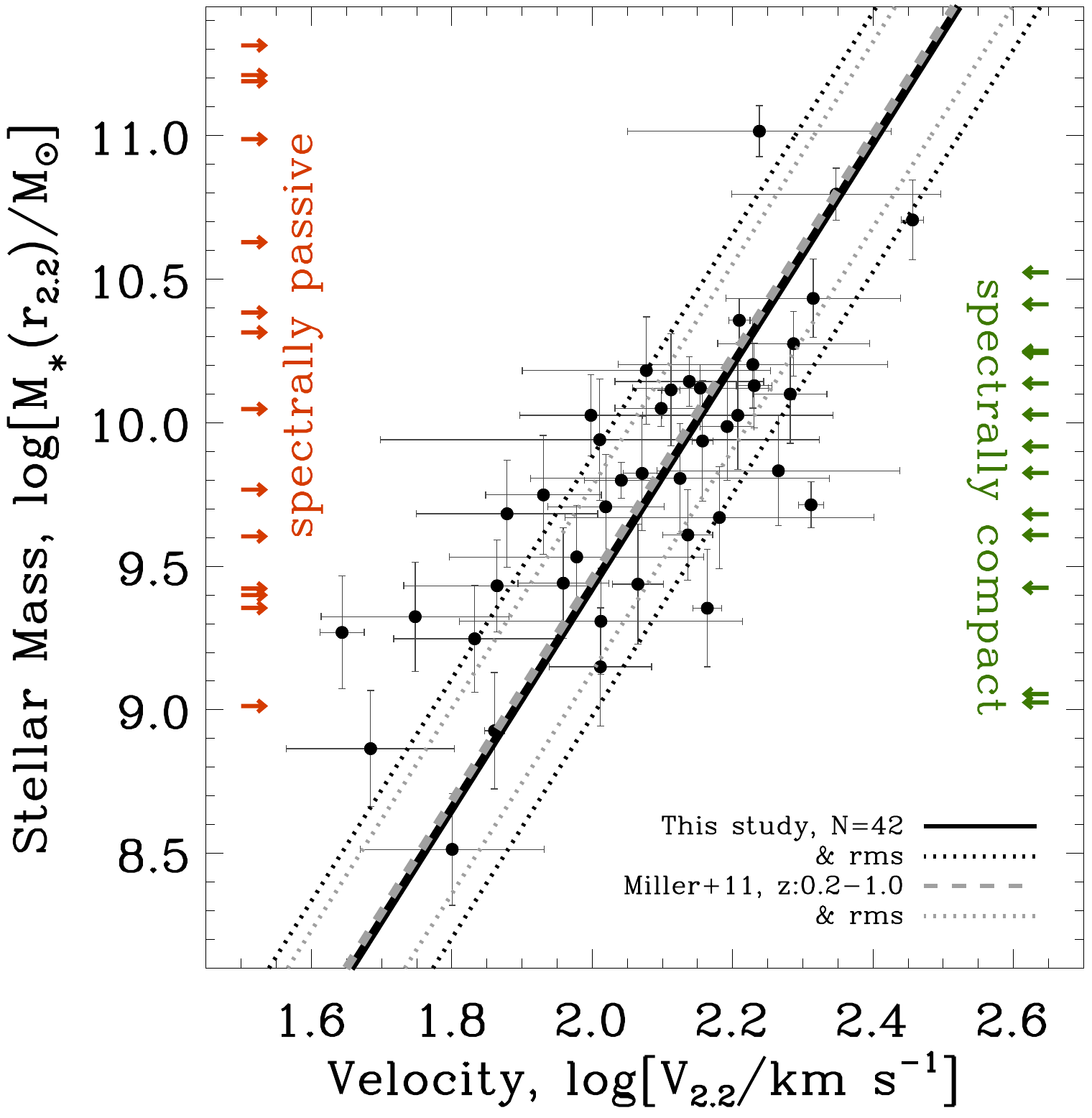}
   \caption{The stellar mass Tully-Fisher relation ($M_{\ast}$-TFR) derived for 42 suitable galaxies in the redshift interval 1$<$$z$$<$1.7. The least squares inverse fit relation is shown as a solid black line and the rms scatter is displayed as a pair of dotted lines. The best-fit $M_{\ast}$-TFR found from $z\sim0.2$ to $z\sim1$ in Paper I is shown as the dashed grey line (which is almost coincident with the results of this study). The rms scatter around the Paper I fit is shown the pair of grey dotted lines. Although the zero point has not changed in the intervening 2 Gyr, the apparent scatter has increased. We also show with arrows the stellar masses of two categories of galaxies for which we cannot extract measurements of $V_{2.2}$ (see \S3.1 for details).}
   \label{fig:tf}
\end{figure}

The primary result of this paper, the stellar mass Tully-Fisher relation ($M_{\ast}$-TFR) for the redshift interval $1<z<1.7$, is shown in Fig.\ref{fig:tf}. As in Paper I, we plot the enclosed stellar mass within $r_{2.2}$ against the de-projected rotational velocity $V_{2.2}$.  To interpret the relationship with respect to our earlier work, we fit a linear regression using a least-squares method incorporating the intrinsic scatter ($\sigma_{int}$) added in quadrature to that of the velocity and accounting for errors in both the ordinate and abscissa. As in Paper I, we focus our analysis on the {\it inverse fit}, assuming velocity as a function of mass. However, following traditional convention, the relation is displayed with $V_{2.2}$ on the x-axis and $M_{\ast}$ on the y-axis. 

Our linear fit can be written as:
\begin{equation}
\log{\left(\frac{{M}_{\ast}}{M_{\odot}}\right)} =  a+b~\log{ \left( \frac{ V_{2.2} }{ \mathrm{km~s}^{-1} } \right) } -\log{\left(\frac{{M}_{0}}{M_{\odot}}\right)},
\end{equation}
where ${M}_{0}$ = ${10}^{10}$$M_{\odot}$. With a slope $b$ =3.869 fixed from that derived for the total sample in Paper I , we find that $a$=1.692$\pm$0.060, $\sigma_{int}$=0.080 dex in log[$V$/km s$^{-1}$], and the total rms is 0.117 dex, which is reasonable considering the median error is 0.102 dex. In stellar mass, this corresponds to $\sigma_{int}$=0.310 dex in log[$M_{\ast}/M_{\odot}$] and an rms of 0.452 dex (where the median error is 0.186 dex).  

While we recover a zero-point for the fixed-slope $M_{\ast}$-TFR similar to that found in Paper I for $0.2<z<1.3$, the scatter in the relation has apparently increased as much as 60\% as compared to the local relation. Part of this increased scatter could be attributed to the more challenging observations undertaken with LRIS. We note, for example, that the additional scatter introduced by using LRIS c.f. DEIMOS as described earlier ($\sim$0.098 dex in log[V/km s${}^{-1}$]), implies our intrinsic scatter could be as low as 0.063 dex. Given the uncertainty in the scatter is about 0.01 dex, this value is not significantly different than that found in Paper I  \citep[0.058 dex,][]{miller2011}. So while it is possible that the relation is less well-defined by a broadening of up to 60\% compared to $z\sim0$, where average scatter is found to be $\sim$0.05 dex \citep[i.e.,][]{pizagn2005, reyes2011}, we cannot rule out the possibility that the tightness of the relation in this study is the same as that found locally.

We also show in Fig.~\ref{fig:tf} the stellar masses for two important subsets of the originally-targeted sample whose resolved dynamics cannot be determined. The first category are the 14 {\it spectrally passive} objects which are clearly some of the most massive galaxies in the sample. Conceivably
these passive galaxies are evolving off the TF relation and transitioning onto the red sequence. On the other hand, the stellar masses of the {\it spectrally compact} objects do not significantly differ from that for main sample displaying suitably extended emission. Most of these galaxies appear to have extended disks in the HST ACS images, although star formation appears to have largely shut off in their outer disks as evidenced by their spectrally compact nature. Conceivably, these galaxies could comprise systems with actively forming bulges, up to a few times $10^{10}$ $M_{\odot}$ in stellar mass.

\begin{figure*}
  \includegraphics[width=3.5in]{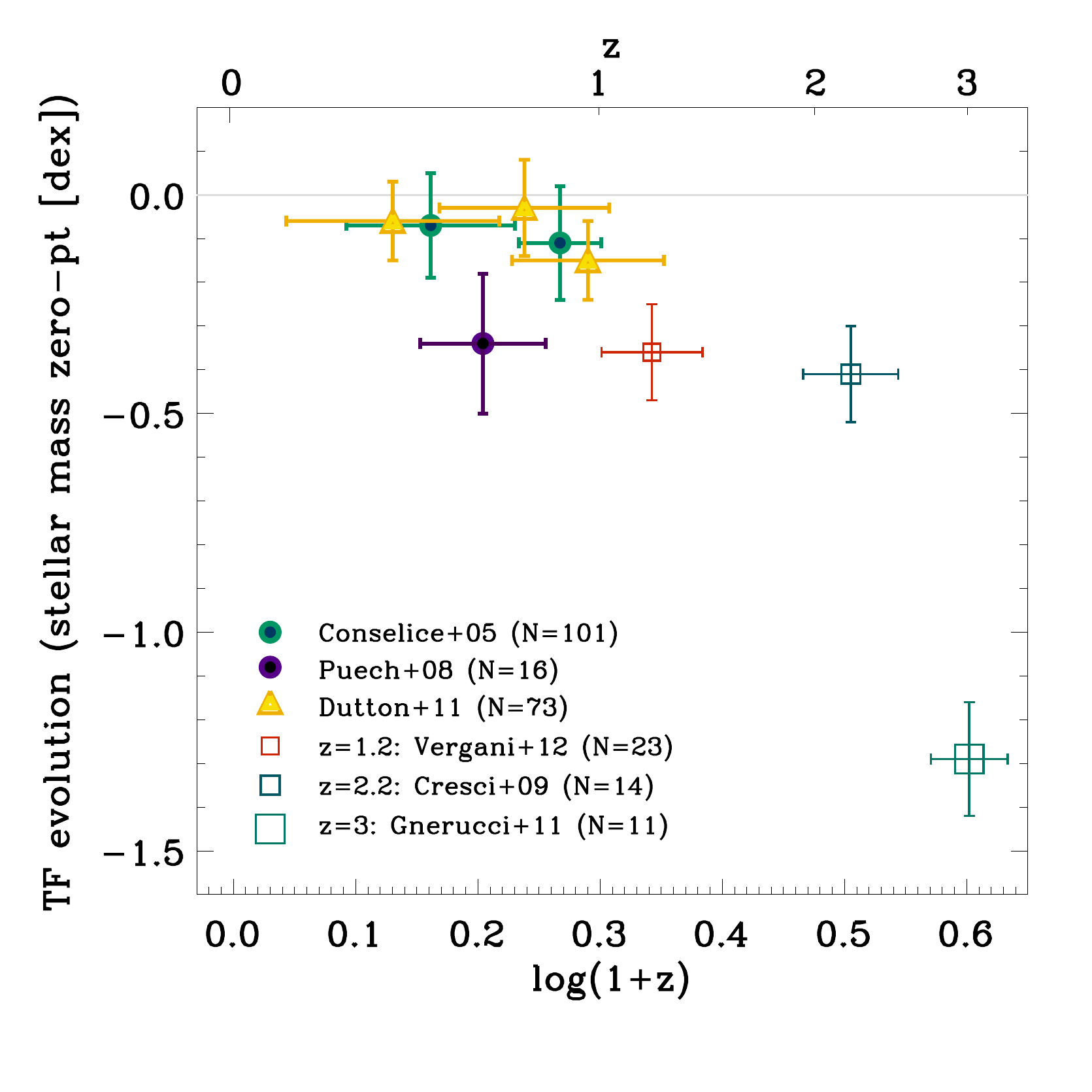}
    \includegraphics[width=3.6in]{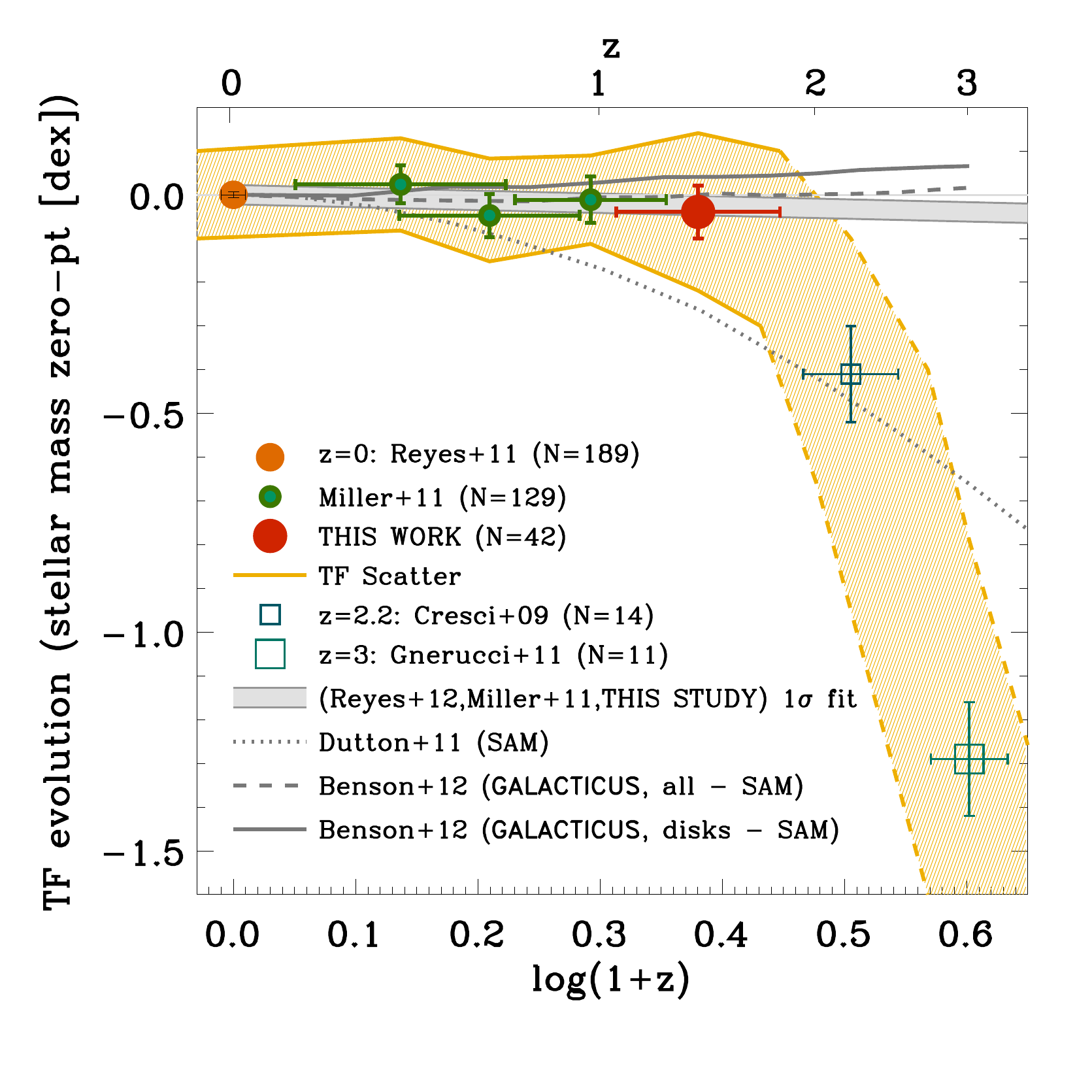}
   \caption{The left panel shows the progress in constraining the evolution of the $M_{\ast}$-TFR zero-point \citep{consel2005,puech2008,dutton2011a} before the present series of papers, with exception of \citet{vergan2012}, which appeared on the arXiv pre-print service while this paper was under review at ApJ. We note that all of the IFU studies \citep{puech2008,vergan2012,cresci2009} lie suggestively offset from the local relation at the same interval in dex, with exception of the highest redshift study done at present \citep{gneruc2011}.  On the right we show the improvement of our results, showing both the redshift dependent $M_{\ast}$-TFR scatter and zero-point, assuming a fixed slope as determined by \citet{reyes2011} at z$\sim$0 (orange circle). We compare the results from this study (red circle) and Paper I results (green circles) to those at higher redshift \citep{cresci2009,gneruc2011}. We also compare semi-analytic models (SAM) from  \citet{dutton2011a} and \citet[{\sc Galacticus},][]{benson2012} For a full discussion of this plot, please see \S\ref{sec:results} and \S\ref{disc}.}
   \label{fig:tf_evo}
\end{figure*}

Finally, to place our $M_{\ast}$-TFR in the context of earlier work, Fig.~\ref{fig:tf_evo} summarizes recent results for the $M_{\ast}$ Tully-Fisher relation over the full redshift range 0$<$$z$$<$3.  Notwithstanding possible selection biases discussed in \S5 it is illustrative to consider this figure. Evolution in the $M_{\ast}$-TFR zero-point can be normalized with respect to the local study of \citet{reyes2011}, calibrated from the parent sample of which came from SDSS and who also adopt the same $V_{2.2}$ parameterization used in this series. To ensure an adequate comparison, we re-fit our data using the \citet{reyes2011} slope,  (which we note is not significantly different from the slope we find). Because \citet{reyes2011} use the total mass rather than that enclosed at $r_{2.2}$, we move their zero-point by -0.187 dex in stellar mass to account for the difference in the enclosed mass as compared to the total stellar mass. We make a further correction to account for the different IMF assumed, since \citet{reyes2011} adopted a \citet{kroupa2002} IMF, which both sets of authors agree accounts for a 0.05 dex decrease compared to our assumed \citet{chabri2003}. In total, this is a -0.137 dex calibration between the \citet{reyes2011} study and our work, resulting in only a marginal change in the $M_{\ast}$-TFR zero-point since $z\simeq1.7$. Formally, the linear regression shown in Fig.~\ref{fig:tf_evo} yields a stellar mass zero-point change of only 0.02$\pm$0.02 dex over the last 9.8 billion years.

We also include results at $z\sim2.2$ and $z\sim3$ from \citet{cresci2009} and \citet{gneruc2011} respectively, which suggest a surprisingly rapid and significant evolution beyond $z\sim1.7$.  However, the average stellar mass in the \citet{cresci2009} sample is 6.3$\times$$10^{10}$ $M_{\odot}$, and range from 0.62--31.6 $\times$$10^{10}$ $M_{\odot}$, whereas our sample average is 1.3$\times$$10^{10}$ $M_{\odot}$, and ranges from 0.05--17.2$\times$$10^{10}$ $M_{\odot}$ (both using a Chabrier IMF).

As a theoretical comparison we present preliminary $M_{\ast}$-TFR evolution results computed using {\sc Galacticus} v0.9.1 r631\footnote{computed with an input parameter file which is available from http://sites.google.com/site/galacticusmodel/downloads/parameter-sets} \citep{benson2012}, which includes various modules for star formation and AGN feedback as well as state-of-the-art theoretical predictions for baryon-halo evolution. Shown in Fig.~\ref{fig:tf_evo} are {\sc Galacticus} models for both disk-dominated galaxies (bulge-to-total mass fraction $<$ 0.3) and all galaxies. In each case, the mean zero-points correspond to galaxies with stellar masses between 0.5 and 2$\times$$10^{10}$ $M_{\odot}$. Interestingly, the results from models of all galaxies shows a similar, non-evolving trend in the $M_{\ast}$-TFR zero-point, whereas the the disk-only sub-sample of the models depart in evolution to the slow side of the observed relation as redshift increases, but never more than 0.1 dex --- even to $z$$\sim$3.  In this figure we also include the semi-analytical model of \citet{dutton2011a}, which is consistent with a sample subset of the results originally presented in \citet[][N=73 from 544 galaxies]{kassin2007}. Their projected evolution departs from the results at $z\sim1$ in this study and Paper I. 

\section{Discussion} \label{disc}

The major goal of this paper was to extend our earlier work in Paper I to higher redshifts to determine the rate at which disk galaxies arrive on the Tully-Fisher relation. Figure ~\ref{fig:tf_evo} presents a puzzling discontinuity in behavior over a relatively narrow period of cosmic history. A key uncertainty in the interpretation of this figure, however, is that not all of the sample  satisfies our criterion of providing a reliable estimate of $V_{2.2}$. Although we have maintained continuity in our selection techniques from Paper I to that in this paper in the redshift interval $z$$\simeq$1.0-1.7, similar criteria of morphological appearance or extended emission lines are difficult to usefully apply to the star-forming galaxies being studied beyond $z\simeq2$. As redshift increases in Fig.~\ref{fig:tf_evo}, some subset of a mass-selected or a star formation rate (SFR)-limited population is likely not included in the TF analysis.

Such a selection bias may be more or less important depending on the physical model we wish to test. If disks grew from chaotic beginnings, where clumps collide at the centers of dark halos \citep{vdberg1996,abraha1997,elmegr2005}, then the TF relation will continue to evolve as the chaotic behavior and mergers \citep{hammer2005} transitions into more regular accretion onto well-formed disks. In this scenario, we could interpret the lack of observed evolution as the result of a bias towards ``mature'' systems at all redshifts.  However, since we expect high SFRs in this violent phase, these less mature systems are more likely to be included in our analysis, unless dust obscuration or possible nuclear concentration of star formation makes extended emission lines difficult to detect (i.e., our spectrally compact and passive targets).

Alternatively if disks form predominantly via smooth accretion over time, they could easily evolve along self-similar dynamical relations, making selection effects much less important in understanding the lack of evolution. How then could we interpret the more drastic offsets implied by Fig.~\ref{fig:tf_evo}? Even with smooth, cold accretion \citep[e.g.,][]{dekel2006}, if galaxies formed stars more vigorously and erratically from $z$$\sim$2--4 than at present times \citep[e.g.,][]{forste2006, genzel2008}, then a tight $M_{\ast}$-TFR would be delayed until disk properties are homogenized across the observed populations--- similarly for a transition from dynamically hot objects to disks which are primarily rotationally supported. Any inhomogeneity, including differences or changes in:
\begin{itemize}
\item rotational versus dispersion support, 
\item baryonic mass-to-light ratio across the disk,
\item baryonic versus dark matter fraction,
\item specific star formation rate, and
\item gas fraction or gas reservoir size and distribution
\end{itemize}
across redshift bins should lead to an evolution in the $M_{\ast}$-TFR. While various combinations of the above characteristics are likely correlated, they may not evolve in tight lock-step. Thus the emergence of a recognizable  $M_{\ast}$-TFR at $z$$\simeq$1.5 is already a result of significance --- even if it only applies to some subset of the star-forming population.

At $z$$\sim$3, \citet{gneruc2011} argue that the nascent TFR has a scatter of over an order of magnitude larger than what we observe at $z$$\simeq$1. Given that the angular momentum properties are shown to be similar to those at $z$$<$1 \citep{bouche2007}, if the sources targeted by \citet{cresci2009} were less efficient star-formers than those studied here, perhaps this could explain the marked stellar mass deficiency indicated in Fig~\ref{fig:tf_evo}. However, the earlier systems display {\it higher} gas fractions and star-formation rates \citep{taccon2010}, and considering that they may have additional dynamical mass supported by higher velocity dispersions only makes this discrepancy worse.  Studies of less massive sources at $z\sim2$ and beyond as well as better constraints on the physical sizes of these components will be valuable in resolving this possible discrepancy. 

\citet{cresci2009} and \citet{gneruc2011} note the decreased rotation-to-dispersion velocity fraction ($V$/$\sigma$) support in galaxies at $z$$\sim$2, as shown also in \citet{forste2009}, as key in explaining the observed $M_{\ast}$-TFR evolution. However this picture, which posits a greater degree of turbulent support in thicker disks at high redshift, should result in {\it less} rotational motion per unit stellar mass at earlier times, in contrast to the trend shown in Fig. ~\ref{fig:tf_evo}. Given the negative $M_{\ast}$ zero-point evolution observed, a reduced $V$/$\sigma$ at higher redshift would thus imply a substantial evolution in the gas fraction. Simultaneous changes in the dark matter fractions, baryonic mass-to-light ratios, and disk geometries may lead to a less severe evolution in gas fraction. Understanding how these characteristics could conspire to keep the $M_{\ast}$-TFR from more dramatic changes can be addressed by more sophisticated tools.

Comparing the relative evolution in our study to that of semi-analytical models can thus provide interesting interpretations as to the predicted interplay between baryons and dark matter over these epochs. We note that the model of \citet{dutton2011a} departs significantly from the observed trends before $z$$\sim$1, whereas the Benson (2012) {\sc Galacticus} models with several more forms of baryonic feedback are a closer match. This suggests that understanding the subtleties in the baryonic physics of galaxies is vitally important to our understanding galaxy assembly via tools like the Tully-Fisher relation. In fact, \citet{dutton2011a} see much less evolution in their baryonic Tully-Fisher relation over the same redshift range, signalling the effect of evolving gas fractions in conjunction with their stellar components. Since the evolution of the dark halo drives the self-similar evolution in disk circular velocity in their model, the evolution present in both their baryonic and $M_{\ast}$-TFRs is being driven by dark matter, similar to the results of \citet{somerv2008a}. However, we find in \citet{miller2011} that baryons may be the dominant driver of dynamics within $r_{2.2}$, and furthermore, the dominance of dark matter would stabilize the disk against the intense star formation which is observed \citep{genzel2008}. Because the role of gas in disk galaxies throughout their formation is not adequately understood across a wide mass range at high redshift, it is likely that further work and exploration of the evolution of the baryonic Tully-Fisher relation will help to explain why we see so little evolution in the $M_{\ast}$-TFR since $z\sim1.7$, especially since we suppose that gas fractions evolve with time \citep{taccon2010}, as well as the dark halo response to baryonic matter.

Our main findings can be summarized as follows:
\begin{enumerate}
\item Using LRIS on Keck I and techniques discussed in Paper I,  we determine reliable rotational velocities using [O II] emission for 42 of 70 targeted star-forming galaxies selected from the redshift range $1.0 \lesssim z  < 1.7$. We test the reliability of our extracted velocities with reference to our data at higher resolution and lower redshift.
\item We find that the stellar mass Tully Fisher relation
for this subsample is well-determined with
up to 60\% increase in scatter compared to the local
relation defined by \citet{pizagn2005}. Considering also the relations
found in Paper I and locally, we find little zero-point shift corresponding to $\Delta M_{\ast}=0.02\pm0.02$ dex from $z=1.7$ to $z=0$.
\item Concerning the 28 galaxies which were not included
in the above analysis, 6 are unresolved both in their broad-band imaging and in their spectra, and 14 are spectrally passive. Although the remaining 8 show extended, resolved broad-band imaging, the emission is insufficiently extended to resolve by LRIS with the extended exposure times of this study. We consider the impact 
of excluding these sources on our derived TF relation and note their properties.
\item The modest evolution seen in the TF relation over $0<z<1.7$
contrasts markedly with results emerging at $z>2$. At face value
it seems there is a dramatic change in the kinematic properties
over a very short period of cosmic history, however a wider range of
sample stellar masses is required at $z>2$ to understand
the validity and significance of this result.
\end{enumerate}

While this paper was under review for publication, a study appeared on arXiv.org [astro-ph] pre-print service, accepted for publication \citep{vergan2012}, which plots 23 galaxies from the MASSIV survey onto a stellar mass TF relation with similar or larger scatter as that found in our relation ($\sigma_{int}$=0.32 or 0.52 and rms=0.48 or 0.72) and little to some evolution in the stellar mass zero-point (-0.05 $\pm$ 0.16 dex or -0.36 $\pm$ 0.11), depending on the \citet{bellde2001} or \citet{pizagn2007} slope adopted, respectively.  While the kinematics data was obtained on the SINFONI IFU, only 5 of these objects were observed using AO (and only 2 in the sub-sample used to plot the stellar mass TF relation). In their TF constraints cited above, they exclude over half of their sample where significant emission was detected, not only when the emission is unresolved, as in our study, but also when their velocity field is not well described by a symmetrically rotating disk, the kinematic position angle differs significantly from the morphologically derived major axis, or when their measured $v/\sigma<1$ (all criteria which are not included in our study). The addition of the modeled velocities in these excluded galaxies adds considerably  to the apparent scatter in the 
\citet{vergan2012} stellar mass Tully-Fisher relation, whereas our inclusion of merger-like systems or galaxies where rotation curve centers appear offset to the HST-derived centers do not appear to result in the same large scatter. This may be due to their use of $V_{max}$, described in this work, Paper I, and e.g. \citet{courte1997} as potentially leading to the underestimation of the true intrinsic maximal rotation velocity because of its dependancy on the distribution of emitting gas in the disk, inconsistent between galaxies in a diverse sample such as theirs. Also we find that assuming an intrinsic emission distribution based on broad-band optical or NIR HST imaging can often lead to systematic offsets in ${\chi}^2$ minimization dynamical modeling, amplified by the scale of seeing and dispersion in the spectral data, as discussed in Paper I of this series. If this effect is significant for slit-based modeling, then it would be just as damaging if not more so in 2-D IFU-based modeling.

If the increased scatter we find in the TF relation in this study from $1<z<1.7$ is truly intrinsic, then we should be able to determine what is different in those systems which have not yet arrived on the TF relation. In future work, we will more closely examine the various characteristics of disk galaxies in our samples from $0.2<z<1.7$:  their estimated mass components, their dynamics, and their bulge-to-disk morphology. The full release of the WFC3 data in the CANDELS fields \citep{koekem2011} will allow us to compare our current bulge-to-disk decompositions to those in rest-frame optical at our highest redshifts observed. For example, it has been postulated \citep[i.e.,][]{bourna2007,jones2010} that the development of a central bulge through inspiraling, unstable clumps is key to stabilizing the disk. Alternatively, morphology could be driven by halo accretion type (hot  versus cold), in which the circular velocity shows little dependency on morphology in the recent work by \citet{sales2011}. By examining the $M_{\ast}$-TF relation as a function of the bulge to disk ratio and morphology, such hypotheses can be directly tested. Furthermore, an in depth comparison between semi-analytic and hydrodynamic models could provide valuable insight on the observed trends.  In doing so, we aim to determine what is driving the emergence of the Tully-Fisher relation we observe by $z\sim1.7$, and potentially, what in the formation processes of $L^{\ast}$ disk galaxies are leading them to largely stabilize by $z\sim1$.


\acknowledgments

SHM thanks the Rhodes Trust, the British Federation of Women Graduates, the sub-department of Astrophysics and New College at the University of Oxford, and the California Institute of Technology for supporting her work. We thank P. Capak for generously allowing us to use his SSA22 photometry catalog. We also thank A. Benson for preliminary results from the {\sc Galacticus} model to compare to our observations, and we thank C. Peng for supplying us with \textsc{Galfit} 3.0. We also thank C. Conroy, A. Dutton, L. Hernquist, J. Gunn, and A. Loeb for helpful discussions on this work. We thank the anonymous referee for improving the quality of this work. The spectroscopic data was secured with the W.M. Keck Observatory on Mauna Kea. We thank the observatory staff for their dedication and support. The authors recognize and acknowledge the cultural role and reverence that the summit of Mauna Kea has always had with the indigenous Hawaiian community, and we are most fortunate to have the opportunity to conduct observations from this mountain.

{\it Facilities:} \facility{Keck I (LRIS)}, \facility{HST (ACS)}.


\bibliographystyle{apj}

\bibliography{mybib}

\appendix
\section{Further Sample Details}

\twocolumngrid

\begin{figure*}
\center
  \includegraphics[width=6.9in]{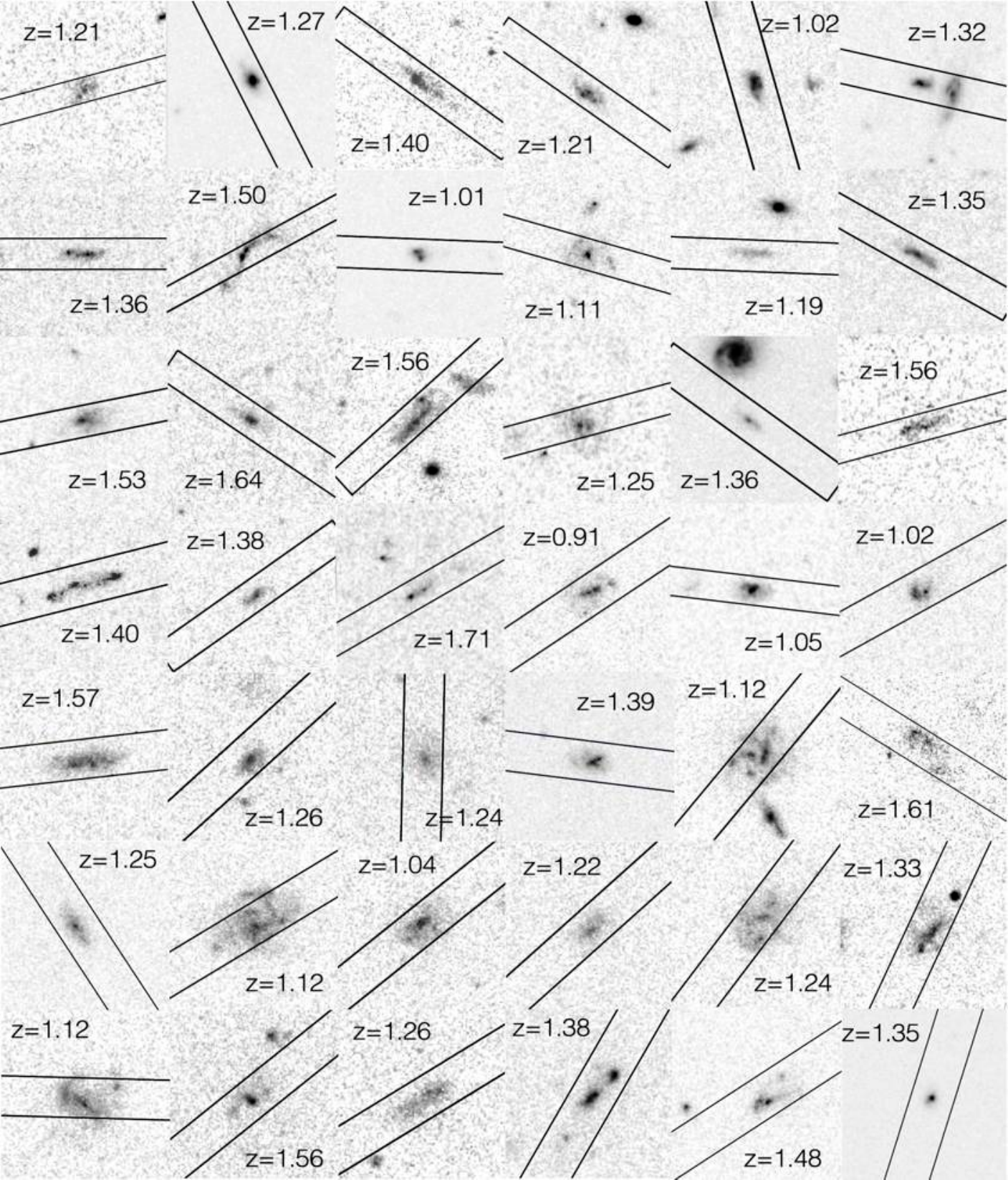}
   \caption{Images of the galaxies in our TF sample, with either the HST F814W or F850LP and their 1 arcsecond slits overlaid. We also include the redshift of each target.}
   \label{fig:samp_mont}
\end{figure*}

Here we more carefully detail the content of our Tully-Fisher sample, suggesting that we have a representative sample of rotationally supported galaxies at $1<z<1.7$, including irregular and peculiar/interacting galaxies.

For our initial selection based on visual morphology, the purpose for excluding unresolved, compact sources is that their 
    dynamical measurements from emission lines are similarly unresolved (and in fact often 
    non-existent). This accounts for an effective removal of about 1/5 of objects, 
    and an additional de-weighting of 1/10 of objects which meet
    the magnitude cut. As seen in the morphological
    montage in Fig. \ref{fig:samp_mont}, the TF sample includes 4 objects (1/10 of sample) that would be morphologically classified as
    early types or nearly unresolved (compact), as well as at least 8 objects (1/5 of sample) which would be considered peculiar, interacting, or undergoing a minor merger. This fraction is consistent with the evolving merger rate fraction at this redshift \citep{cassat2005,consel2008,bridge2010}, whether or not the initial visual morphological selection was biased toward mergers.  
    
    A different picture is presented in works by \citet{flores2006,kassin2007}, and \citet{puech2008} which cite or suggest higher interacting and merging fractions (at least 45\% of their samples with complex kinematics, with an additional 28\% considered to be perturbed rotators in the latter work),  which is used to explain increased scatter in the Tully-Fisher relation in these works. While it is well known in local studies that peculiar and interacting galaxies add some scatter to the idealized Tully-Fisher relation of a homogenous population \citep[e.g.,][]{tutui1997,kannap2002,mendes2003}, this level of scatter is similar to what we find in our work. Rather \citet{flores2006,kassin2007}, and \citet{puech2008} argue for the important contribution of peculiar and interacting galaxies on the larger scatter in their relations. Although methods for identifying mergers vary in more complete samples at the same redshifts, e.g., \citet{kartal2007,jogee2009} \citep[it is not trivial to compare these due to different timescales of merger signatures,][]{consel2009}, we nevertheless believe that measurement uncertainty from decreased signal-to-noise and resolution contributes to increased scatter in previous intermediate-z studies, rather than an increased scatter arising primarily from disturbed or complex kinematics  \citep[see also, e.g.,][]{kannap2004,miller2011}. We have attempted in this work and in \citet{miller2011} to clarify this important difference with extended integrations on spectra and a well-tested dynamical modeling method for deriving intrinsic rotation velocities.
    
 The three objects of overlap between our Paper I sample 
    and that of \citet{flores2006} and \citet{puech2008} are all in the most
    kinematically complex classification of the \citet{flores2006}
    scheme, so this suggests that we are not removing objects which 
    may contribute maximally to the scatter seen in the TF relation,
    according to \citet{puech2008}. Instead, as mentioned before, we see a similar amount of scatter as local studies
    which are inclusive to irregular, perturbed, and peculiar 
    galaxies, \citet[i.e.,][]{reyes2011}, which has a sample constructed to be representative based on SDSS.

We find in Paper 1 for $0.2<z<1.3$ galaxies that with only 1 hour of data, 1/6 of the sample would have been in our spectrally compact category and 1/3 of the sample would have been considered passive in emission, i.e., not detected at all by DEIMOS. So over half of the sample that we include in our Tully-Fisher relations would be prematurely excluded in samples that make a significant emission cut based on $\sim$1 hour redshift surveys from similarly sensitive instruments/telescopes as DEIMOS/Keck, such as the [OII] emission criteria used to construct the MASSIV survey sample \citep{contin2011,vergan2012}.

While in Paper I we implement a slitlet position angle offset (from major axis) velocity correction, significant in only 10 (8\%) of objects, no such corrections were found necessary in this study (see Fig. \ref{fig:samp_mont}). 


\end{document}